\documentclass[a4paper,fleqn,usenatbib]{mnras}

\usepackage[T1]{fontenc}
\usepackage{ae,aecompl}

\usepackage{graphicx}	

\usepackage{amsmath}	
\usepackage{amssymb}	

\title[Excess entropy and Non-gravitational energy in galaxy clusters] 
{Little evidence for entropy and energy excess beyond $r_{500}$ - An end to ICM preheating?
}
\author[Iqbal et~al.]
{Asif Iqbal$^{1}$\thanks{
asifiqbal@kashmiruniversity.net}, Subhabrata Majumdar$^{2}$\thanks{
subha@tifr.res.in}, Biman B. Nath$^{3}$\thanks{
biman@rri.res.in}, Stefano Ettori$^{4,5}$\thanks{stefano.ettori@oabo.inaf.it}, 
\newauthor Dominique Eckert$^{6}$\thanks{Dominique.Eckert@unige.ch} and Manzoor A. Malik$^{1}$\thanks{
mmalik@kashmiruniversity.ac.in}\\
  $^{1}$Department of Physics, University of Kashmir, Hazratbal, Srinagar, J\&K, 190011, India \\
  $^{2}$Tata Institute of Fundamental Research, 1 Homi Bhabha Road, Mumbai, 400005, India\\
  $^{3}$Raman Research Institute, Sadashiva Nagar, Bangalore, 560080, India\\
  $^{4}$INAF, Osservatorio Astronomico di Bologna, Via Ranzani 1, I-40127, Bologna, Italy \\
  $^{5}$INFN, Sezione di Bologna, viale Berti Pichat 6/2, 40127, Bologna, Italy\\
  $^{6}$Astronomy Department, University of Geneva 16, ch. d'Ecogia, CH-1290 Versoix, Switzerland
}
\pubyear{2016}

\begin{document}
\label{firstpage}
\pagerange{\pageref{firstpage}--\pageref{lastpage}}
\maketitle

\begin{abstract}
Non-gravitational feedback affects the nature of the intra-cluster medium (ICM). 
X-ray cooling of the ICM and {\it in situ} energy feedback from AGN's and SNe as well as {\it preheating} of the gas at epochs preceding the formation of clusters are proposed mechanisms for such feedback.
While cooling and AGN feedbacks are dominant in cluster cores, the signatures of a preheated ICM are expected to be present even at large radii. To estimate the degree of preheating, with minimum confusion from AGN feedback/cooling, 
we study the excess entropy and  non-gravitational energy profiles upto $r_{200}$  for a sample of 17 galaxy clusters using  joint data sets of {\it Planck} SZ pressure and {\it ROSAT/PSPC} gas density profiles. 
The canonical value of preheating entropy floor of $\gtrsim 300$ keV cm$^2$, needed in order to match cluster scalings, is ruled out at $\approx 3\sigma$.
We also show that the feedback energy of 1 keV/particle  is ruled out at 5.2$\sigma$ beyond $r_{500}$. Our analysis takes both non-thermal pressure and clumping into account which can be important in outer regions.
Our results based on the direct probe of the ICM in the outermost regions do not support any significant preheating.   

\end{abstract}



\section{INTRODUCTION}
Galaxy clusters are the largest and most massive virialized objects in the universe, which make them ideal probes of the
large scale structure of the universe and, hence, of cosmological parameters that govern the growth of structures  (see \cite{1} and references therein). 
However, in order to obtain robust estimates of these parameters, using X-ray techniques, one requires precise knowledge about the evolution of galaxy clusters with redshift and the thermodynamical properties of intracluster medium (ICM). 
In the simplest case, where one considers pure gravitational collapse, cluster scaling relations are expected to follow self-similarity \citep{Kaiser1986, Sereno2015}.  X-ray scaling relations have been widely used to test the strength of correlations between cluster properties and to probe the extent of self-similarity of clusters \citep{Morandi2007}. These observations show departure from self-similarity;
for example, the luminosity-temperature ($L_x-T$) relation for self-similar models predict a
shallower slope ($L_x\propto T^{~2}$) than observed ($L_x\propto T^{~3}$). Similarly, Sunyaev-Zel'dovich (SZ) scaling relations also show similar departure \citep{Holderb2001}. 
 
Such departures point towards the  importance of complex non-gravitational  processes over and above the shock heating of the ICM. 
The first idea aimed at explaining departure from self-similar scaling relations is that of {\it preheating}, first proposed by  \cite{Kaiser1991} and later extended by others \citep{Evrard1991,Babul2002}.  In this scenario, the cluster forms from an already preheated and enriched gas due to feedback processes (such as galactic winds or AGN) heating up the surrounding gas at high redshifts. 
Preheating models require constant entropy level of  $\gtrsim 300$ keV cm$^2$ in order to explain the  break in the self-similarity scaling relations \citep{Tozzi2001,Babul2002,McCarthy2002}.
In terms of ICM energetics, this typically translates into feedback energy of $\sim1$ keV per particle \citep{Tozzi2001,Pipino2002,Finoguenov2003}. 
However, there is an ambiguity in defining preheating energy/particle since it depends on the density at which gas is heated (less dense gas requires smaller energy to raise it to
a particular entropic state). Therefore, preheating is best expressed in terms of entropy.

Although, early preheating models could describe the scaling relations in clusters, it had drawbacks with regard to details. For example, 
these models predicted isentropic cores particularly in the low mass clusters \citep{Ponman2003} and an excess of entropy in the  outskirts of
the clusters \citep{Voit2003} which are not consistent with observations. The idea of preheating has endured and has found resurgence in recent times (see \cite{Pfrommer2012,Lu2015} and references therein). \cite{Pfrommer2012} suggested time dependent entropy injection due to TeV blazars which provide uniform heat at $z\sim3.5$  peaking near $z\sim1$ and subsequent formation of CC (NCC) clusters by early forming groups (late forming groups) while \cite{Lu2015}  explored preventative scenario of feedback in which the circum-halo medium is heated to finite entropy. 

In contrast to preheating, there can also be {\it in situ} effects such as injection of energy feedback from AGN, radiative cooling, supernovae and 
star formation, influencing the thermal structure of ICM \citep{Roychowdhury2005,10,Eckert2013a}. 
There is growing evidence that AGN feedback mechanism provides a major source of heating for the ICM \citep{McNamara2007,Fabian2012,Chaudhuri2013} in the cluster cores.
Outside cluster cores, however, the estimates of entropy floor and feedback energy  (particularly in massive clusters) are more reflective of preheating of gas since (i) the effect of central sources is unlikely to be significant and (ii) the loss of energy through radiation is negligible.

It is worth noting that irrespective of the nature of feedback, the thermodynamic history of the ICM is fully encoded in the entropy of the ICM. The 
ICM entropy profile is defined as\footnote{Thermodynamic definition of specific entropy being $S=\ln K^{3/2}+$ constant} 
 $K(r)=k_BTn_e(r)^{-2/3}$, where $k_B$ is the Boltzmann constant. 
Non-radiative AMR/SPH simulations, which encodes only gravitational/shock heating, predict entropy profiles of the form $K(r) \propto r^{1.1}$ \citep{Voit2005}. 
Apart from slightly larger normalization, it has been found that there is significantly higher (flatter) core entropy in  AMR case as a result of the hydrodynamical processes that are resolved in the code
(e.g. shocks and mixing motions) \citep{Mitchell2009,Vazza2011,Power2013}. On the other hand, observations find deviations from the predicted entropy profile at small radii \citep{10,Eckert2013a} as well as large radii \citep{Eckert2013a,Su2015}.

A meaningful comparison of recent observations with
theoretically expected entropy profiles can thus be used to determine the nature and degree of feedback. This idea was developed and used recently by \cite{Chaudhuri2012,Chaudhuri2013} who estimated the non-gravitational energy deposition profile in the cluster cores. They 
 compared benchmark non-radiative AMR/SPH  entropy profiles \citep{Voit2005} with observed entropy profiles for the {\it REXCESS} sample of 31 clusters \citep{10}
and found the excess mean energy per particle to be $\sim 1.6 -2.7$ keV up to $r_{500}$. 
Further, they  showed that the excess energy  is strongly correlated with AGN feedback in cluster cores \citep{Chaudhuri2013}.

In the present study, we extend their work by going beyond $r_{500}$ and estimate entropy floor and feedback energetics at large cluster radii. 
The effect of clumping and non-thermal pressure, especially at large radii, has been shown to be important \citep{Eckert2015, Battaglia2015,Shaw2010,Shi2015} and we incorporate both in our analysis.

We study the joint data set of {\it Planck} SZ pressure profiles and {\it ROSAT} gas density profiles of 17 clusters \citep{Eckert2012,Planck2013a} to estimate entropy profiles up to $r_{200}$ and beyond\footnote{We have left out cluster ``A2163'' from \cite{Eckert2013a,Eckert2013b} in this work as its estimated feedback profile was found hugely different from others. This cluster is in the perturbed state and presumably out of hydrostatic equilibrium \citep{	
Soucail2012}.}. As detailed in \cite{Eckert2013a}, we use the parametric profiles which are obtained by fitting a functional form to projected emission-measure density and {\it Planck} SZ pressure data \citep{Vikhlinin2006, Nagai2007}\footnote{www.isdc.unige.ch/$\sim$deckert/newsite/Dominique\_Eckerts\_Hom\\epage.html.}.  The parametric profiles have less cluster-to-cluster scatter and errors; however, they are consistent with the non-parametric deprojected profiles. Below $0.2~r_{500}$, the resolution of both {\it Planck} and {\it ROSAT} is insufficient to obtain reliable constraints.

In the last 25 years since its proposal, the evidence for-or-against preheating has been mainly circumstantial. In this {\it Letter}, we show that a direct estimate of entropy floor and non-gravitational energy in the outer regions is insignificant enough so as to rule out preheating scenarios. Throughout this work, we will assume ($\Omega_m$, $\Omega_{\Lambda}$, $h_0$) = (0.3, 0.7, 0.7).
\section{ANALYSIS}
\subsection{Cluster modeling}
The  total hydrostatic mass profile $M(r)$ of the galaxy clusters is given by
$M(r)\, =\,-\frac{r^2}{G\rho_g(r)}\,\frac{dP_g(r)}{dr}$,
where  $\rho_g$ and $P_g$ are the parametric forms of the density and thermal pressure of the ICM respectively \citep{Eckert2013a,Planck2013a}.
The radii $r_{500}$ and $r_{200}$  are  obtained by first interpolating the $M(r)$ profile and then iteratively solving \footnote{$\Delta$ is defined such that $r_{\Delta}$ is the radius out to which the mean matter density is $\Delta \rho_c$, where $\rho_c=3H^2(z)/8\pi G$ 
being critical density of the universe at redshift $z$.}  for $m_{\Delta}\,=\,(4/3)\,r^3_{\Delta}\,\Delta \,\rho_c(z)$.
The virial radius, $r_{vir}(M_{vir},z)$, is calculated with spherical collapse model
$ r_{vir}=\left[\frac{M_{vir}}{4\pi/3\Delta_c(z)\rho_c(z)}\right]^{1/3}$
where $\Delta_c(z)=18\pi^2+82(\Omega_m(z)-1)-39(\Omega_m(z)-1)^2$. If required, virial radius is obtained by linear  extrapolation of mass profile in  logarithmic space.

Since the ``actual'' total mass is also partially supported by non-thermal pressure, we model the non-thermal pressure fraction using the form given in \cite{Shaw2010},
\begin{equation}
 P_{nt}(r,z)\,=\,f(r,z)\,P_{tot}\,=\,\frac{f(r,z)}{1+f(r,z)}\,P_g(r),
 \label{eqn2:eq2}
\end{equation}
where $P_{tot}$ is total gas pressure, $f(r,z)=a(z)\left(\frac{r}{r_{500}}\right)^{n_{nt}}$,   $a(z)=a_0(1+z)^{\beta}$ with $a_0=0.18\pm0.06$, $\beta=0.5$ and $n_{nt}=0.8\pm0.25$ \citep{Shaw2010}. We also study the effect of different non-thermal pressure fraction by varying $a_0$.
For our sample, the fiducial $P_{nt}$ is  $\sim50\%$ of the thermal gas pressure, $P_{g}$, around $r_{vir}$ and corresponds to a mass difference of order $20\%$ at $r_{500}$. 
This is in good agreement with simulations/theoretical predictions \citep{Shi2015}. The value of $r_{500}$ obtained from the resultant mass profiles are consistent with the \cite{Planck2011}. 
With the addition of the non-thermal pressure, the value of $r_{500}$ typically increases by $50-150$ Kpc; however, this difference is degenerate with the value of the normalization of $P_{nt}$.

\subsection{Initial entropy profile}
Models of the formation of the large scale structure, where gas is shock heated as it  falls in the cluster dark matter potential well,  predict that the gas entropy $K_{th}(r)$ has
a power-law behavior with radius outside of cluster cores. 
For non-radiative AMR simulations, \cite{Voit2005} entropy profile  is well described in the range $(0.2-1)~r_{200}$ by
\begin{equation}
 \frac{K_{th}(r)}{K_{200}}=1.41\left(\frac{r}{r_{200}}\right)^{1.1},
 \label{eq:voit1}
\end{equation}
 plus a flatter core below 0.2$\,r_{200}$ with $K_{200}$=$144 \left(\frac{m_{200}}{10^{14}M_\odot}\right)^{2/3} \left(\frac{1}{f_b}\right)^{2/3} h(z)^{-2/3} \mbox{ keV cm}^2$. We fix $f_b$=$0.156$ from the recent {\it Planck} results \citep{Planck2013b}.
It has been found that the entropy profiles after taking cooling into account differ with Eq.~\ref{eq:voit1} significant only up 300 Kpc for $10^{15}$ solar mass cluster \citep{McCarthy2008} which corresponds  to $\approx0.2~r_{500}\approx0.1~ m_g/m_{g,500}$ for our sample. 

The hydrostatic equation, now including both thermal and non-thermal pressure, can be rewritten in terms of the entropy as
\begin{equation}
\frac{d(P_g+ P_{nt})}{dr}\, =\,-\left( \frac{P_g}{K_{th}} \right )^{3/5}  m_p \mu_e^{2/5} \,\mu^{3/5} \, \frac{ G M_{tot}( <r ) }{ r^2} ,
\label{H:he2}
\end{equation}
where $M_{tot}$ is the total mass which is equated to $M_{thermal} + M_{non-thermal}$. For  boundary condition, we fix the gas fraction ($f_g$) to be $0.9f_b$ at virial radius \citep{Crain2007}. 
Initial profiles for density and temperature are found using Eqs.~\ref{eq:voit1} \& \ref{H:he2}.

Recently, both simulations and observations have found significant clumping beyond $r_{500}$, which by definition is measured as $C$=<$\rho_g^2$>/<$\rho_g$>$^2$,  \citep{Eckert2013a,Eckert2015,Battaglia2015}. 
 \cite{Eckert2015} found azimuthal median is a good tracer of the true
3D density (clumping factor) and showed from both  hydrodynamical simulations and synthetic simulations that their method recovered the true 3D density profiles with  deviations less than 10\% at all radii.
They found that the average  $\sqrt C=1.25$ at $r_{200}$, consistent with the numerical simulations. Since clumping in the ICM is a plausible reason for the observed flattening of the entropy profiles in the 
outer regions, we estimate the observed entropy profiles by incorporating clumping using the recent parametric form of the clumping profile given in section 4.1 of \cite{Eckert2015}.

 \begin{figure}
\centering
\includegraphics[width=8 cm]{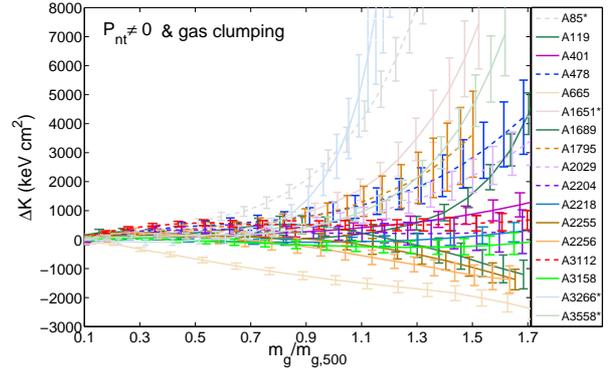}
\caption{The excess entropy $\Delta K$  as a function of $m_g/m_{g,500}$ for all clusters. Solid and dashed lines represent NCC and CC clusters respectively.
Four clusters marked with $*$ have large value of $\Delta K$ ($>4000$ keV cm$^2$) in outer regions and are not included in sub sample (see Tab. \ref{table1}). The error bars are given at 1$\sigma$ level.
}
\label{fig:AmrdeltaEind}
\end{figure} 
\subsection{Estimates of total feedback energy}
To estimate the feedback thermal energy, we need to relate the entropy change (i.e., $\Delta K=K_{obs}-K_{th}$) with change in energy. Considering isobaric approximation,  thermal energy change per unit mass is given by
$\Delta Q={kT_{obs}  \over (1-{1 \over \gamma})\mu m_p} {  \beta ^{2/3} (\beta -1)\, \over (\beta^{5/3}-1)} {\Delta K \over K_{obs}} $ (see \cite{Chaudhuri2012} for details), where $\beta=T_{obs}/T_{th}$ and $\gamma=5/3$. 
Most importantly, in order to take into account the redistribution of gas mass due to the feedback one should compare entropy profiles for the same enclosed gas mass (i.e., $\Delta K(m_g)$) instead at the same radii ($\Delta K(r)$) as commonly done in the literature) \citep{Li2010,Nath2011,Chaudhuri2012,Chaudhuri2013}. 
The corresponding mechanical feedback energy per particle ``$\Delta E_{\rm ICM}$'' can be written  
in terms of change in thermal and potential energies as
\begin{equation}
\Delta E_{\rm ICM}=\mu m_p\Delta Q+G\mu m_p\left(\frac {M_{tot}(r_{th})}{r_{th}}-\frac{M_{tot}(r_{obs})}{r_{obs}}\right), 
\end{equation}
where $r_{th}$ and $r_{obs}$ are theoretical and observed radii respectively enclosing the same gas mass.
The total amount of feedback energy available in the ICM is  $ E_{\rm ICM} \,= \,\int \frac{\Delta E_{\rm ICM}}{\mu m_p}  \, dm_g \,.$

Since clusters lose energy due to X-ray cooling, we estimate total feedback energy deposited in the ICM by adding this lost energy to $E_{\rm ICM}$; thus $\Delta E_{\rm feedback}= \Delta E_{\rm ICM}+ \Delta L_{bol}\,t_{age}$,
where $\Delta L_{bol}$ is the bolometric luminosity in a given gas shell which is obtained by using the approximate cooling function  $\Lambda_{N}$ given by \cite{Tozzi2001} and $t_{age}$ is the average age of 
the cluster which we have approximated to be $5$ Gyr based on the results of \cite{Smith2008}. Finally, we estimate the {\it mean} non-gravitational energy per particle,  <$\Delta E$>, from total energy divided by the total number of particles in the ICM (i.e $\frac{M_{gas,obs}}{\mu m_p}$). 

In the rest of the paper, we refer to the case where the energy lost due to cooling is not added to energy estimated from entropy differences as {\it final (after cooling)}, 
i.e., $\Delta E_{\rm ICM}$. In contrast,  where the energy lost due to cooling is also added  is referred to as
{\it initial (before cooling)}, i.e., $\Delta E_{\rm feedback}$. {\it The latter represents the non-gravitational energy/particle required to heat the gas in a collapsed system from
the initial theoretical model to the observed state. However, if the change in configuration is solely due to preheating of gas much before the collapse of system then the amount of energy required would be less than $\Delta E_{\rm feedback}$ \citep{McCarthy2008}. 
This implies that $\Delta E_{\rm feedback}$ represents upper an limit on the preheating energy}.

\begin{figure}
\centering
\includegraphics[width=8 cm]{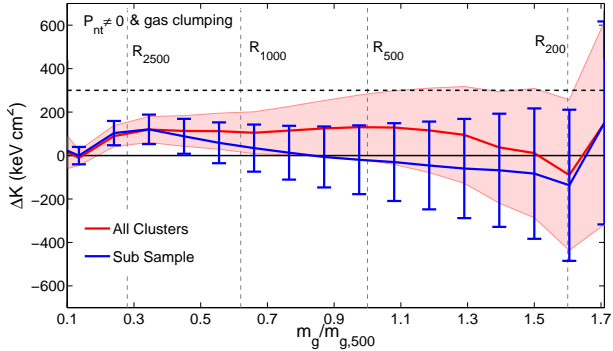}
\caption{The excess entropy $\Delta K_{\rm feedback}$  as a function of $m_g/m_{g,500}$. The thick red  line shows weighted average profile with 3$\sigma$ error  for the entire sample.
Blue line represents average profile for sub sample. The vertical dashed lines show the radius of the mean profile for different overdensities. The horizontal  black line shows zero entropy and the dashed black line is for $\Delta K_{\rm feedback} = 300$ keV cm$^2$, indicative of preheating.}
\label{fig:AmrdeltaE}
\end{figure} 
\section{Results and Discussion}
\subsection{Feedback beyond $r_{500}$}
\begin{figure}
\centering
\includegraphics[width=8 cm]{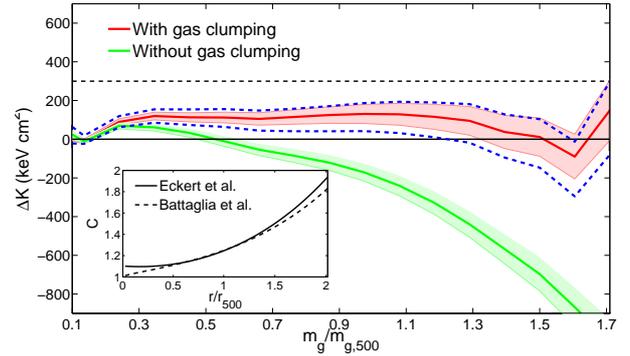}
\caption{
This plot shows the effect on the  $\Delta K_{\rm feedback}$ profile by introducing clumping factor using \citet{Eckert2015} best fit.
The shaded region shows 1$\sigma$ error. The region enclosed by two dashed blue lines  show the  1$\sigma$ error band after accounting for clumping errors (15\% of the clumping profile).  The inset shows comparison of  \citet{Eckert2015} and \citet{Battaglia2015} clumping profiles for the average case.}
\label{amrdeltEwithclumping}
\end{figure}

Once the individual profiles are found, we study the mean properties of the sample. The magnitude and profiles of $\Delta K$ and $\Delta E$, estimated following the method laid down, provide clue to the feedback  on the ICM. 
In Fig.~\ref{fig:AmrdeltaEind}, we see the weighted average \citep{Louis1991} $\Delta K$ profile is close to $100$ keV cm$^2$ for most of the cluster region. 
There are four cluster marked with * in Fig.~\ref{fig:AmrdeltaEind}, which are not included in sub sample for which $\Delta K$ profiles have comparatively large value (and hence large positive change in thermal energy) in outer regions. However, after accounting  for the change in potential
energy along with change in thermal energy, the  
$\Delta E$ profiles for these clusters become close to zero (or even negative). Moreover, for the sub-sample, the $\Delta K=0$  is always consistent at 1$\sigma$ beyond $r_{1000}$. Fig.~\ref{amrdeltEwithclumping} shows $\Delta K$ with and without including clumping in calculations.

\begin{figure}
\centering
\includegraphics[width=8cm]{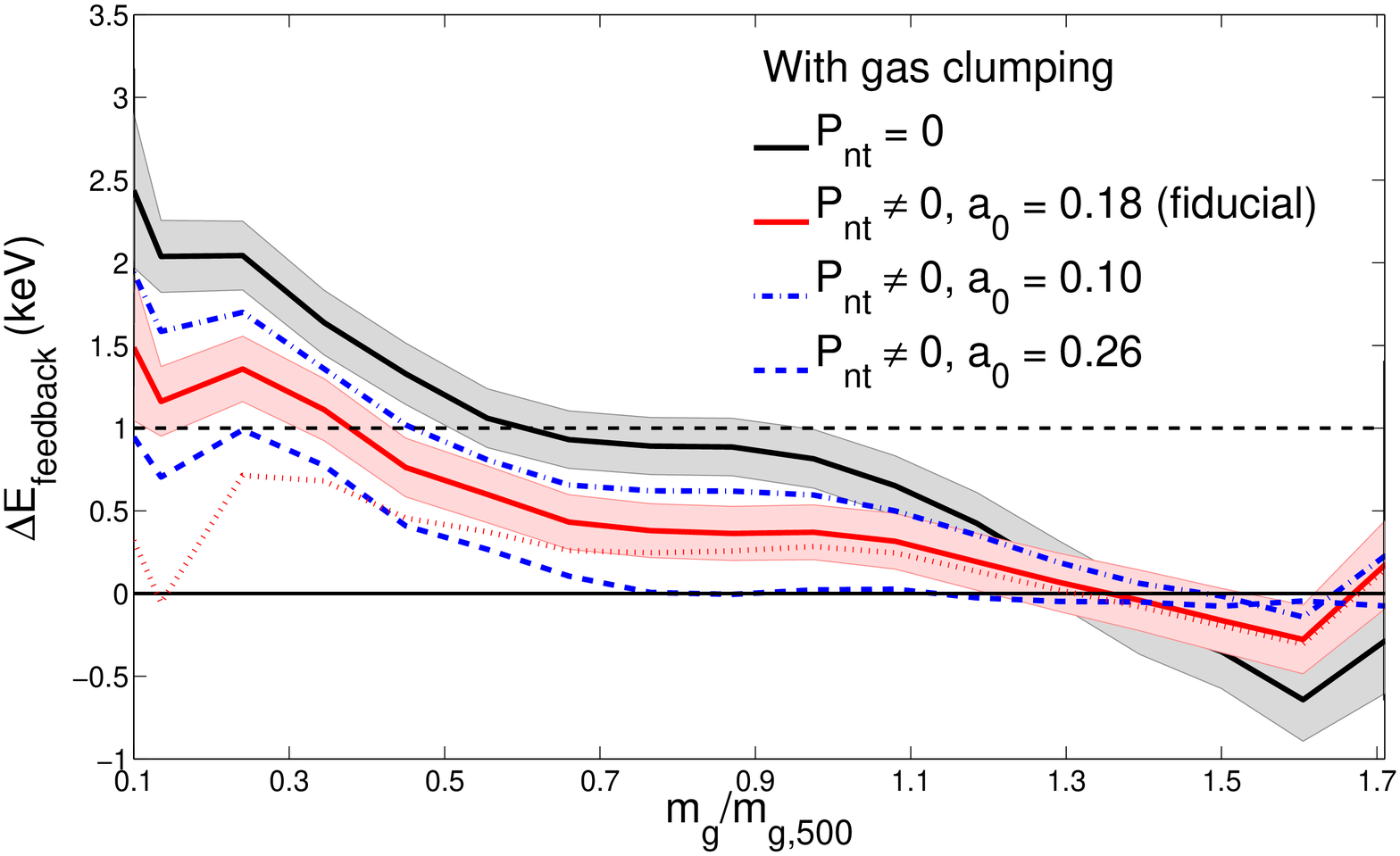}
\caption{
 This plot shows the  $\Delta E_{feedbck}$ for different normalization $a_0$ of the non-thermal pressure with larger value of $a_0$ giving larger non-thermal pressure (see Eq.~\ref{eqn2:eq2}). We show the 1$\sigma$ error bands for the fiducial case (i.e, $a_0 = 0.18$, red band) and the purely thermal case (i.e, $a_0 = 0$, gray band). We  also show the average profile without adding the energy lost due to cooling (i.e $\Delta E_{ICM}$) with dotted red line for the  fiducial case. For meaningful comparison, we have scaled the x-axis of all the cases with the same $m_{g,500}$ as that of fiducial case.}
\label{amrdeltEnor}
\end{figure}

In Fig.~\ref{amrdeltEnor}, we show the corresponding  average  $\Delta E_{\rm feedback}$ (solid red line) for the full sample and compare it with the average of $\Delta E_{\rm ICM}$ (dotted red line).
These are indistinguishable beyond $r \sim r_{500}$ since, unlike in the inner region (as explored in \cite{Chaudhuri2013}), cooling plays sub-dominant 
role beyond $r_{500}$. There is clear evidence of the feedback up to  $\approx r_{500}$ with the feedback peaking centrally (also found by \cite{Chaudhuri2013}). However, the average $\Delta E$ profile 
is close to zero beyond $r_{500}$. Since, more than $70\%$  of the cluster volume lies between $r_{500}-r_{200}$, one can confidently claim insufficient or complete lack of feedback over most of the cluster volume.  

\begin{table*}
 \caption{Average feedback energy per ICM particle (in keV) after including non-thermal pressure and clumping.}
 \begin{tabular}{@{}lcccc}
  \hline
 &\multicolumn{2}{c}{{\it final} average feedback energy/particle }&\multicolumn{2}{c}{{\it initial} average feedback energy/particle}\\
\cline{1-5}
Sample         & $(0.2-1)~r_{500}$  &$r_{500}-r_{200}$ &$(0.2-1)~r_{500}$ &$r_{500}-r_{200}$\\
  \hline
Full Sample    &$0.35\pm0.17$ ($0.34\pm0.17$)    & $0.03\pm0.18$ ($0.11\pm0.17$) & $0.72\pm0.17$  ($0.72\pm0.17$) &  $0.05\pm0.18$ ($0.14\pm0.17$)    \\
\hline
Sub Sample     &$0.60\pm0.21$ ($0.60\pm0.21$)    & $ 0.11\pm0.18$ ($0.11\pm0.18$)  & $1.00\pm0.21$ ($1.00\pm0.21$)  &  $0.13\pm0.18$ ($0.13\pm0.18$)     \\
\hline
 \end{tabular}

Columns (2) \& (3): Average energy per particle in the range $(0.2-1)~r_{500}$ and $r_{500}-r_{200}$ respectively without taking into account energy lost due to cooling (i.e., {\it final} feedback energy ``$\Delta E_{\rm ICM}$''). 
Columns (4) \& (5):  Average energy per particle in the range $(0.2-1)~r_{500}$ and $r_{500}-r_{200}$ respectively after taking into account energy lost due to cooling (i.e.,  {\it initial} energy ``$\Delta E_{\rm feedback}$'').
The numbers in brackets show the average energy per particle for boundary condition $f_g=0.9f_b$ at the last observed radius instead at virial radius. Error bars are given at $1\sigma$ level. Clearly, there is little evidence of feedback energy beyond $r_{500}$ for all cases. 
\label{table1}
\end{table*}
\subsection{Discussion}
It is now amply clear that both non-thermal pressure and clumping are important at large radii. The addition of non-thermal pressure increases the initial entropy profile ``$K_{\rm th}(m_g)$'' due to the increase in the normalized 
$K_{200}$. This in turn leads to the decrease in $\Delta K$ and hence $\Delta E$ 
(see \cite{Iqbal2016b} for details). Considering the clumpiness in gas density (and assuming that no fluctuations exist in temperature distribution), however, results in increase in the observed entropy and hence increase in
the $\Delta E$.  The importance of clumping ($K\sim C^{5/6}$) is highlighted in Fig.~\ref{amrdeltEwithclumping}, where we show the average $\Delta K$ profile before and after correcting for the clumping bias. 
While the estimated entropy excess is unrealistically negative when no correction is applied, it attains a postive value close to zero when the effect of clumping is taken into account following the parametrization of \cite{Eckert2015}. 
Note that this determination is consistent with the expectation of numerical simulations (\cite{Battaglia2015}, see the inset of Fig.~ \ref{amrdeltEwithclumping}).
We find, {\it preheating value of entropy floor $\geq 300$ keV cm$^2$ is ruled out at $3\sigma$ for full sample and at 4.2$\sigma$ for sub sample.}

To study the impact of non-thermal pressure on the estimate of non-gravitational
energy, we  show the $\Delta E$ profiles for the pure thermal case along with the non-thermal case with three different normalization $(a_0=0.10, 0.18, 0.26)$ 
in Fig.~\ref{amrdeltEnor}. These correspond to mass differences of $\sim(10\%, 20\%, 30\%)$ at $r_{500}$ for the average profile. The mean 
excess energy is still far below 1 keV/particle and consistent with zero beyond a specific radius which depends on the choice of $a_0$. 
However, neglecting non-thermal pressure overestimates the feedback energy, though still staying less than 1 keV in the outer regions. 

Finally, we list the average energy/particle  in Tab.~\ref{table1}.  
We find, {\it beyond $r_{500}$, the  $\Delta E_{feedback}$=1 keV/particle, is ruled out at 5.2$\sigma$ for the full sample, and by 4.8$\sigma$ for sub sample}. Since,  $\Delta E_{feedback}$ is roughly the upper limit of preheating energy/particle, this  in turn rules out 
preheating scenarios which require 1 keV/particle to explain the break in scaling relations. At regions below $r_{500}$, $\Delta E = 1$ keV/particle 
is allowed within 3$\sigma$.  It may be also noted from the table that our results are insensitive to the choice of the boundary conditions, particularly for the sub sample. Thus, our constraint on extra heating refers to the inner regions ($<r_{1000}$) only, which strongly corroborate with the results of \cite{Gaspar2014}. Our results can be compared to the value obtained by \cite{Chaudhuri2013} who studied the regions inside 
the core ($r < 0.3 r_{500}$) and obtained $1.7\pm 0.9$ keV/particle which they showed to be strongly correlated to the central AGN feedback \footnote{Note, that \cite{Chaudhuri2013} did not consider $P_{nt}$ or clumping.}.  
The feedback energy left in the ICM is much lower for the entire radial range with cooling influencing the average energy per particle mainly in the range $0.2-1~r_{500}$.
\section{Conclusions}
Our analysis shows that the estimated entropy excess and energy input corresponding to this excess of the ICM is much less than required by preheating scenarios to explain the break in scaling relations. 
While the feedback energy estimates rely on some assumptions (isobaric and cooling energy approximations) and refer to energy deposition after the collapse of cluster, the constraints on the $\Delta K$ shows that 
preheating scenarios that require $\Delta K$ more than 300 keV cm$^2$ can be ruled out.
This result holds good whether or not the effects of non-thermal pressure and clumping are taken into account.
At large radii, the effect of central sources is unlikely to be significant \citep{Hahn2015}, and the loss of energy through radiation is also 
negligible. While some previous workers have cast doubts on the simple preheating scenario arguing that no single value of energy input can explain
the observations \citep{Younger2007}, one can in principle construct variations in the scenario \citep{Fang2008} in order to explain observations 
that are dominated by processes in the inner regions. However, our analysis directly probes the entropy floor and energetics of the cluster gas at the outermost regions 
and shows that any significant preheating that can manifest as a property of the ICM is absent.
\section*{Acknowledgements}
This work was supported by SERB (DST) Project Grant No. SR/S2/HEP-29/2012. AI would like to thank TIFR, Mumbai and RRI, Banglore for hospitality. 
The project started with discussions when two of the authors (SM \& SE) were sharing an office at the Munich Institute for Astro- and Particle 
Physics (MIAPP) in 2015. SM would like to thank Nick Kaiser and Dick Bond for comments. DTP preprint no. TIFR/TH/16-12. 
The authors would like to thank the anonymous referees for useful comments that helped in improving the clarity of the manuscript.
\footnotesize{
    \bibliographystyle{mn2e}

}
\bsp	
\label{lastpage}
\end{document}